\documentclass[10pt,twoside,twocolumn]{IEEEtran}
\usepackage{graphicx,cite,calc,epstopdf}
\usepackage{tikz}
\usetikzlibrary{shapes,arrows}
\usepackage{rotating}
\usepackage{amsmath,amssymb,amsfonts}
\usepackage{algorithmic}
\usepackage{textcomp}
\ifCLASSINFOpdf
\else
\fi

\begin{document}
%
\title{Efficient Sensing of Correlated Spatiotemporal Signals: A Stochastic Gradient Approach}


\author{Hadi~Alasti,~\IEEEmembership{Senior Member,~IEEE}
\thanks{H. Alasti is with the School of Polytechnic, Purdue University Fort Wayne (PFW), Fort Wayne, IN, 46805 USA (e-mail: halasti@ieee.org).}
}

\maketitle

\begin{abstract}
A significantly low cost and tractable progressive learning approach is proposed and discussed for efficient spatiotemporal monitoring of a completely unknown, two dimensional correlated signal distribution in localized wireless sensor field. The spatial distribution is compressed into a number of its contour lines and only those sensors that their sensor observations are in a $\Delta$ margin of the contour levels are reporting to the information fusion center (IFC). The proposed algorithm progressively finds the model parameters in iterations, by using extrapolation in curve fitting, and stochastic gradient method for spatial monitoring. The IFC tracks the signal variations using these parameters, over time. The monitoring performance and the cost of the proposed algorithm are discussed, in this letter.
\end{abstract}
\IEEEpeerreviewmaketitle

\section{Introduction}
\IEEEPARstart{S}{patiotemporal} monitoring (STM) of correlated signal distributions has been explored for smart environment monitoring applications such as monitoring the temperature of hot island \cite{hot_island_monitoring}, gas density monitoring \cite{Vuran_1,Kramer}, monitoring the city air pollution \cite{air_quality_1,IGARSS,Dutta_Sensors_16,Urban_ieee_sens}, and in other applications such as medical image processing~\cite{Costilla-Reyes}, remote sensing~\cite{Gokaraju}, etc. In STM, depends on the complexity of the signal distribution, the sensors may need to process and transmit massive amount of information, over time. Limited sensor resources, such as energy, available bandwidth for communications and computation capability mandates that the reporting sensors and their observations are correctly selected. Energy conservation techniques such as sensor selection \cite{Zhang_Selection}, compressed sensing \cite{Hadi_MISS}, compress and forward \cite{compress_forward}, statistical filtering \cite{Hadi-Thesis,Alasti_FOWANC}, etc. have been employed, so far to let the wireless sensor network's energy last longer.

\begin{figure}[t]
		\centering
		\includegraphics[width=3.1in,angle=0]{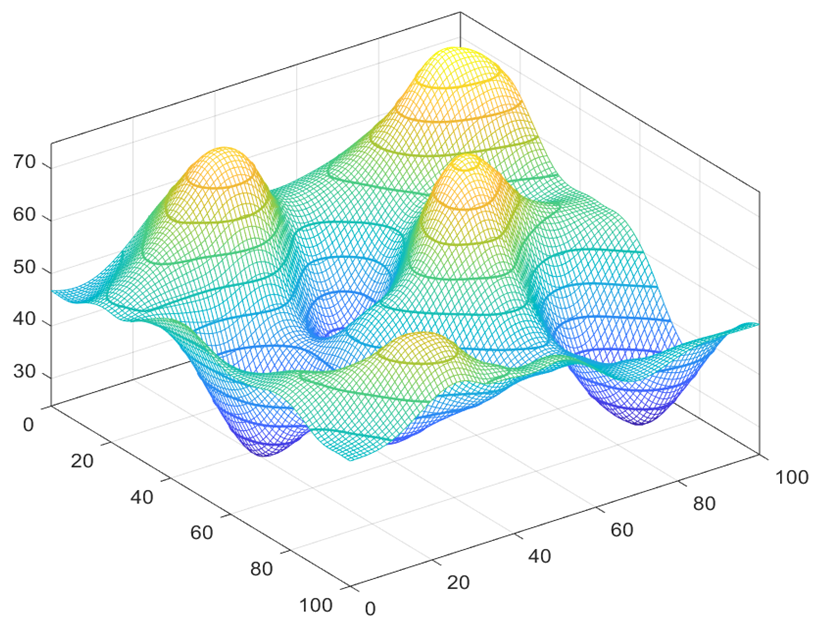}\\
		\caption{Modeling a spatial distribution using its contour lines at $M$ contour levels $\{\ell_i\}_{i=1}^M$.}
		\label{fig: 1}
\end{figure}

This article presents an algorithm, which allows the information fusion center (IFC) to efficiently monitor a spatial distribution such as $g(x,y;t)$ over time, by calling a small subset of the sensor observations through an iterative process. The algorithm models the spatial distribution with $M$ number of its contour lines, as it is illustrated in Fig.~\ref{fig: 1}, at levels $\{\ell_i\}_{i=1}^M$, where $M$ and the contour levels $\ell_i, i = 1,2,\cdots M$ are initially unknown. The proper number of contour lines and their levels are calculated in the process of spatial monitoring based on the proposed approach in~\cite{Hadi_MISS}. During the iterations, the algorithm improves the spatial distribution estimation and its cost progressively using a stochastic gradient method. The signal strength range estimation is improved during extrapolation in curve fitting. We introduce and discuss the performance of a stochastic gradient algorithm, which results in significant saving in the number of transmissions. In this article, we assume that the wireless sensors are localized, which means the IFC knows their coordinate. The assumed uncertainties in the signal distribution are the range of the signal strength; the statistical characteristics of the signal; and its spatial, spectral and temporal attributes. 
This letter is organized as follows. In the next section, the related works are reviewed. Then in section III, the STM algorithm is detailed. The performance and cost of the proposed algorithm is discussed, in section IV.  
\section{Related Works}
Modeling spatial distributions such as images using their contours lines has taken the researchers' attention and been practiced for years. Modeling using contours is a non-uniform sampling technique~\cite{Marvasti} and a sub-category of level crossing sampling (LCS) that has been addressed as compressive sensing~\cite{Guan, Candes, Donoho}. It is worthwhile to mention that LCS has special application interests in sampling sparse signals~\cite{Hadi-IET-WSS, Hadi-IGI} and compressed sensing~\cite{Li}. One importance of LCS is its potential to spontaneously sample related to the bandwidth.

Contour line detection in wireless sensor networks, which is the first step in modeling the spatial distribution has been addressed in several researches, including~\cite{Hadi_MISS,Alasti_FOWANC,Hadi-GreenCom,Liao_1,Liao_2,Liao_3,Liao_4,Armstrong,Gandhi,Suri,Sarkar}. Most of these mentioned researches addressed to distributed-contour-detection that is based on collaboration among sensors for detection of the contour lines. In this letter, we propose a cost efficient centralized algorithm based on the approach proposed in~\cite{Hadi_MISS}.

Spatial modeling of signal distributions using contour lines has been addressed in~\cite{Hadi_MISS,Alasti_FOWANC,Lian}. Modeling the spatial distribution with uniformly spaced contour levels and tracking their variation using time-series analysis in sensors was studied in~\cite{Lian}. Using non-uniformly spaced contour lines was reported first in~\cite{Alasti_FOWANC}. They assumed the probability density function (\textit{pdf}) of the signal strength and used \textit{Lloyd-Max} method to calculate the optimal/sub-optimal contour levels. An iterative algorithm was proposed in~\cite{Hadi_MISS} to extract the \textit{pdf} of the signal strength with low cost in spatial monitoring of the signal distribution.

Spatiotemporal modeling using machine learning approaches has been reported in several researches, including~\cite{Costilla-Reyes,Gokaraju,Dai}. In most of these approaches, neural networks algorithms, genetics algorithms, stochastic gradient descent algorithms, etc. are employed.

In this letter, we propose an algorithm that progressively improves the performance of the model. After convergence of the spatial monitoring, the IFC iteratively updates the changes based on spatial model parameters.
\section{STM Algorithm}
\begin{figure}[b]
		\centering
		\includegraphics[width=3.0in,angle=0]{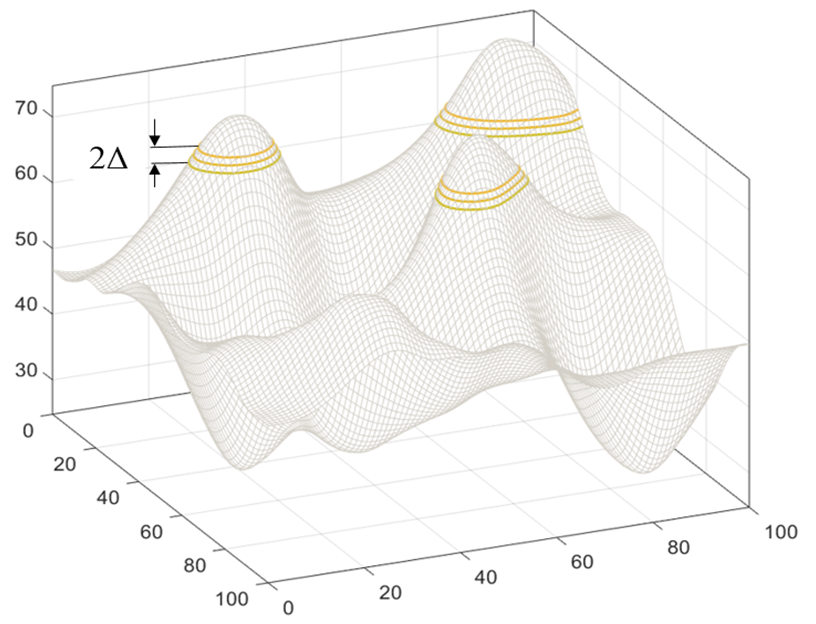}\\
		\caption{Only a subset of sensors that their local sensor observations $S_k, \forall k$ are in $\Delta$ margin of each contour level are reporting to the IFC.}
		\label{fig: 2}
\end{figure}

The proposed algorithm in \cite{Hadi_MISS} introduced an iterative approach for spatial monitoring of an unknown two-dimensional signal distribution from limited sensor observations. It models the spatial distribution with $M$ contour lines at levels $\{\ell_i\}_{i=1}^M$, as it is shown in Fig.~\ref{fig: 1}. Then, a selected subset of sensors, whose their observations $S_k , \forall k$ are in a given $\Delta$ margin of each contour level, $\ell_i - \Delta \leq S_k \leq \ell_i - \Delta, i=1,2,\cdots,M , \forall k$ as it is shown in Fig.~\ref{fig: 2}, report their observations to the IFC. The sensor observations are fed into bi-harmonic spline interpolator ~\cite{Sandwell} to reconstruct an improved approximation of the spatial distribution in each iteration. The proposed algorithm in~\cite{Hadi_MISS} introduces an optimal/sub-optimal solution, however it is costly, because i) a large number of sensors need to report to the IFC, during the spatial monitoring as well as in the temporal monitoring processes; ii) there is no measure to select $\Delta$, where large $\Delta$ results in costly monitoring. Moreover, the signal strength range is unknown and needs to be discovered.

As the spatial distribution is modeled with its $M$ contours lines, similar to one dimensional signals, and to reduce the modeling error, the contour levels (comparable with quantization levels) are selected non-uniformly and according to \emph{Lloyd-Max} algorithm~\cite{Sayood} for minimum reconstruction error. These $M$ levels are calculated at IFC, using equations (\ref{equ: 1}) and (\ref{equ: 2}). 
\begin{equation}
\ell_i = \frac{\int_{y_i}^{y_{i+1}}xf_g(x)dx}{\int_{y_i}^{y_{i+1}}f_g(x)dx}, ~~~~i=1,2,\cdots, M
\label{equ: 1}
\end{equation}
where $y_i$ is as follows and $f_g(x)$ is \textit{pdf} of the signal strength.
\begin{equation}
y_i = \frac{\ell_i + \ell_{i-1}}{2}, ~~~~ i = 1, 2, \cdots, M-1
\label{equ: 2}
\end{equation}
\subsection{The Proposed Spatial Monitoring Algorithm}
The spatial monitoring algorithm introduced in \cite{Hadi_MISS}, iteratively estimates the probability density function \textit{pdf}, however it is costly yet and the signal strength range and the contour line margin $\Delta$ are also unknown. In this section, two mechanisms are employed that results in autonomous estimation of the signal strength range and the contour level margin $\Delta$.\\
In this article we use \textit{norm-1} for evaluation of signal estimation error, according to~(\ref{equ: 3}).
\begin{equation}
Error_n = \sum_{i=1}^P\sum_{j=1}^Q\frac{\left|\tilde{g}_n(x_i,~y_j)-\tilde{g}_{n-1}(x_i,~y_j)\right|}{P\times Q}
\label{equ: 3}
\end{equation}
In (\ref{equ: 3}), $\tilde{g}_n(x_i,y_j)$ is the reconstruction of the spatial distribution in the $n$th iteration of the spatial monitoring algorithm, at the $P \times Q$ grid points of the sensor field.\\
\emph{- Signal strength range estimation:}\\
The IFC, sends a query to a few (at least two) arbitrary sensors and requests for their sensor observations from the field's signal distribution. The minimum and the maximum of these readings provide an initial guess for the signal strength range, $(L_{min} , L_{max})$. 
Then the IFC initiates the spatial monitoring algorithm by querying the sensor field and asks for the report of those sensors that their observations are in a $\Delta$ margin of equally spaced levels $\{\ell_i\}_{i=1}^M$, where $L_{min} < \ell_i < L_{max}, \forall i$. The initial $M$ must be small, for instance 3 to reduce the cumulative cost. The initial $\Delta$ can be selected arbitrarily, for instance $\Delta = (\ell_1+\ell_2)/2$. Upon receiving the sensor observations from the field, they are passed to the bi-harmonic spline interpolator that has been introduced in~\cite{Sandwell}. The interpolator's output signal range introduces the new signal strength range for the next iteration steps. In the next iterations, based on the level selection scheme, it can be either equally spaced, or non-uniformly spaced as~\textit{Lloyd-Max} describes in (\ref{equ: 1}) and (\ref{equ: 2}). In each iteration step, $M$ is incremented ($M \leftarrow M+1$) and then the IFC introduces a new set of $M$ levels to the sensor field. Experimental results show that in a few iteration steps the algorithm spans the actual signal strength range.\\
\emph{- The stochastic gradient adaptation:}\\
As the number of the contour lines of the modeled spatial distribution increases, its modeling error statistically decreases, as discussed in~\cite{Hadi_MISS}. Here, we use the decreasing trend of the modeling error to adjust the $\Delta$ in each iteration step. The increase of error is interpreted as insufficiency of the number of reporting sensors, which is proportional with $\Delta$, and vice versa. Accordingly, we incorporate $\Delta$ with the gradient of the reconstruction error function as it is described in (\ref{equ: 4}):
\begin{equation}
	\Delta_k = \Delta_{k-1} (1 + \mu (Error_{k-1} - Error_{k-2}))
	\label{equ: 4}
\end{equation}
In equation (\ref{equ: 4}), $\mu$ is a real and positive value. To improve the learning speed, to stabilize the adaptation process and to eliminate the unknown parameter $\mu$, we normalize the error difference (the error gradient) in equation (\ref{equ: 4}) to the error magnitude, where it stabilizes the adaptation of $\Delta$, similar to~\cite{Bershad}. As such, equation (\ref{equ: 4}) turns into (\ref{equ: 5}).
\begin{equation}
	\Delta_k = \Delta_{k-1} (1 + \frac{Error_{k-1} - Error_{k-2}}{Error_{k-1} + Error_{k-2}})
	\label{equ: 5}
\end{equation}
The $\Delta$ adaptation is called stochastic gradient descent, because in updating $\Delta$ in (\ref{equ: 5}), it follows the slope of the variation of error, and in (\ref{equ: 5}) it uses the normalized gradient.\\
\emph{- Noise reduction using moving average:}\\
Here we assume zero-mean Gaussian noise with average power of $\sigma^2$ in sensor observations. To alleviate the effective power of noise, a moving average filter with $m$ taps is used, where it drops the effective average noise power to $\sigma^2/m$. It is assumed that during the moving average filtering the spatial distribution does not tangibly change. 
\begin{table}[b]
\noindent\textit{Summary of the spatial monitoring algorithm}\\
\line(1,0){245}
\begin{enumerate}
	\item The IFC queries a few selected sensors for an initial guess for the signal strength range $(L_{min} , L_{max})$.
	\item The IFC queries the sensor field for the sensor observations $S_k$ at equally spaced contour levels $\{\ell_i\}_{i=1}^M$, within $(L_{min} , L_{max})$.
	\item Take $\Delta = (\ell_2-\ell_1)/2$ as initial $\Delta$ value
	\item The sensors, in which their sensor observation $S_k$ is in the range of $\ell_i - \Delta \leq S_k \leq \ell_i + \Delta$ respond to the query and send their ID along with $S_k$, for any valid $\ell_i$.
	\item The IFC uses \emph{bi-harmonic spline interpolation} to make a new approximate reconstruction of the signal distribution. Then, it estimates the mean absolute error.
	\item The IFC updates the signal strength range $(L_{min} , L_{max})$.
	\item The IFC uses \emph{Kolmogrov-Smirnov} test (or similar methods) to find the \emph{pdf} of the signal distribution~\cite{Conover}.
	\item The IFC updates the new $\Delta$ according to equation (5).
	\item The number of levels is incremented for one unit, i.e. $M \leftarrow M+1$.
	\item The IFC either uses \emph{Lloyd-Max} (equations 1 and 2), or equally spaced levels to estimate the new $M$ contour levels.
	\item The IFC queries the sensor field for the sensor observations with new $M$ contour levels $\{\ell_i\}_{i=1}^{M}$.
	\item Repeat from Step (4), if required.
\end{enumerate}
\line(1,0){245}
\end{table}
\subsection{Temporal monitoring (spatial tracking)}
Upon convergence of the spatial monitoring algorithm, the IFC uses the final parameters such as the number of required contour lines $M$, and the most recent contour level margin $\Delta$ for temporal monitoring. As the IFC does not need to discover these parameters, the temporal monitoring  has much lower cost (the number of reporting sensors). The temporal monitoring is repeated periodically. In each period, the IFC queries the sensor field by sending the number the most recent contour levels $\{\ell_i\}_{i=1}^M$ and the most recent $\Delta$. Upon receiving the query reply from those sensor observations that are in $\Delta$ margin of these levels, the IFC reconstructs the signal distribution, finds the new signal range and then calculates the most recent $M$ contour levels for the next iteration step. In temporal monitoring iterations $M$ does not increment.
\section{Performance Evaluation}
For performance evaluation of the proposed algorithm, here we use synthetic signals. In generation of synthetic spatial distributions, we used diffusion process model for its simplicity and its flexibility in analytically shifting the local Gaussian terms for temporal evaluation. Diffusion process model has been introduced in~\cite{Jindal_1, Jindal_2}. In this study, the spatial distribution is formed according to (\ref{equ: 6}), where $G(m_{x_i},m_{y_i},\sigma)$ is a two dimensional jointly Gaussian distribution with respectively mean values of $m_{x_i}$ and $m_{y_i}$ and standard deviation of $\sigma$. The coefficients $a_i$ and $b_j$ are positive random values in a known range.
\begin{equation}
g(x,y) = \sum_{i=1}^{N_1} a_i G(m_{x_i},m_{y_i},\sigma_a) + \sum_{j=1}^{N_2} b_j G(\acute{m}_{x_j},\acute{m}_{y_j},\sigma_b)
\label{equ: 6}
\end{equation}
In~(\ref{equ: 6}), the mean values $x_i$ and $y_i$ are selected randomly in the interval (0, 100). The values of $\sigma_a$ and $\sigma_b$ that are forming the spectral characteristics of the spatial distribution are 10 and 3, respectively. $N_1$ and $N_2$ are 150. The spatial signal distribution is spread in rectangular area of $100\times100$. In this area, 5000 wireless sensors are randomly scattered with uniform distribution. The standard deviation of zero-mean noise is assumed 0.3, after moving average filtering. For temporal variation, only the Gaussian terms with standard deviation $\sigma_b = 10 $ in (\ref{equ: 6}) are moved towards horizontal direction.

In performance evaluation, the signal distribution has been modeled using its contour lines with 3 different schemes: i) \textsl{(U-SG)}- $\Delta$ adapted, uniformly spaced contour levels, when the signal strength range is unknown; ii) \textsl{(LM-fixed)}- fixed $\Delta$, contour lines based on \textit{Lloyd-Max} when the range and the \textit{pdf} of the signal strength are known; iii) \textsl{(LM-SG)}- $\Delta$ adapted contour lines based on \textit{Lloyd-Max}, when the range and the \textit{pdf} of the signal strength are unknown. In this performance evaluation, we used \emph{MATLAB}. The related codes are available in \cite{STM_codes}.
\begin{figure}[b]
		\centering
		\includegraphics[width=3.1in,angle=0]{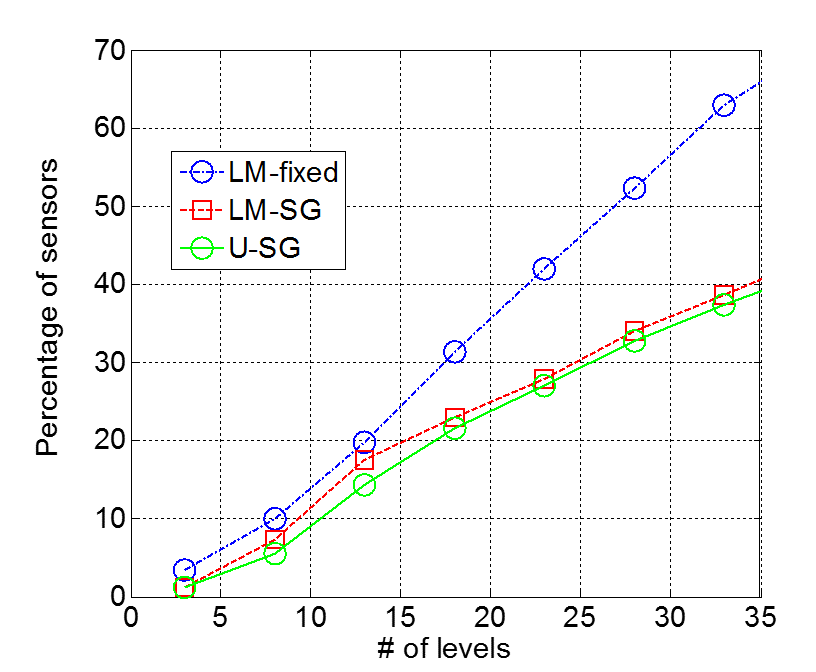}\\
		\caption{The cumulative cost of the spatial algorithm for the three level types: \textsl{LM-fixed}, \textsl{LM-SG} and \textsl{U-SG}, after averaging.}
		\label{fig: 3}
\end{figure}
\begin{figure}[t]
		\centering
		\includegraphics[width=3.1in,angle=0]{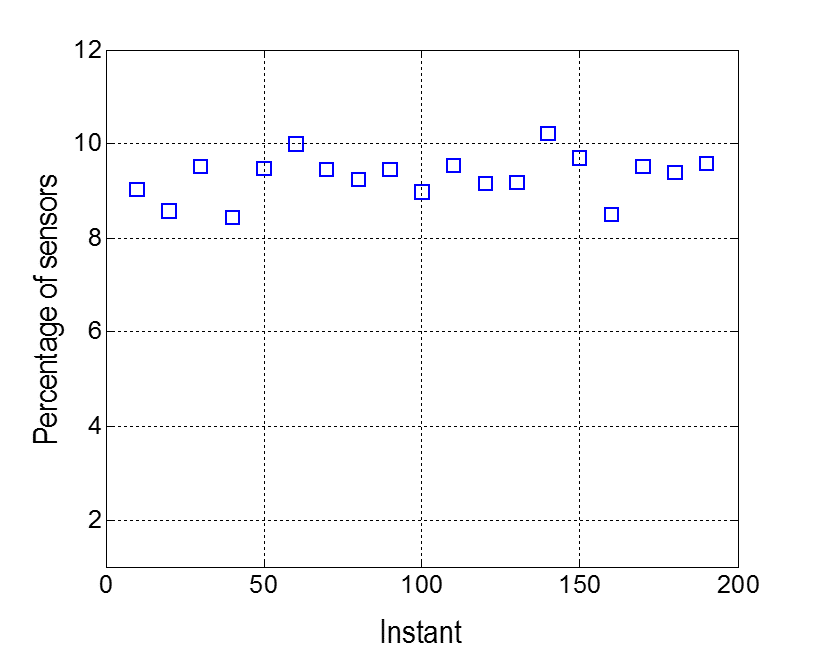}\\
		\caption{The cost of temporal monitoring algorithm for \textsl{LM-SG}.}
		\label{fig: 4}
\end{figure}
\begin{figure}[h]
		\centering
		\includegraphics[width=3.1in,angle=0]{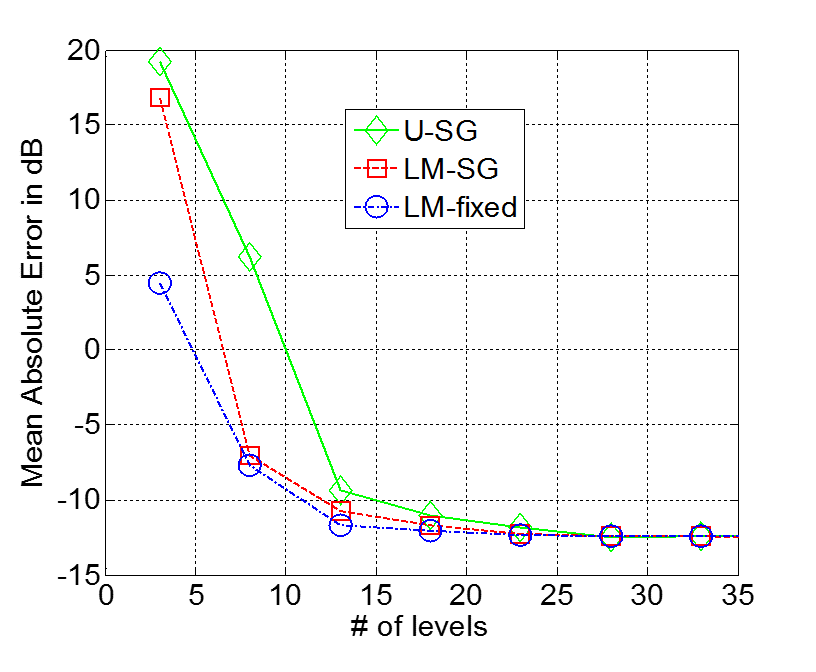}\\
		\caption{The mean reconstruction error of the spatial algorithm for the three level types: \textsl{LM-fixed}, \textsl{LM-SG} and \textsl{U-SG}, after averaging.}
		\label{fig: 5}
\end{figure}
\begin{figure}[h!]
		\centering
		\includegraphics[width=3.1in,angle=0]{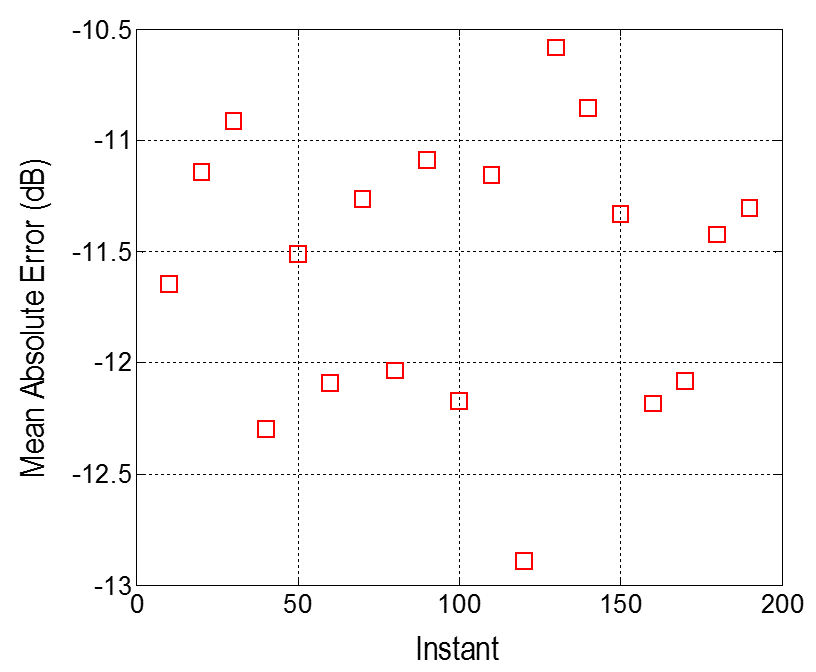}\\
		\caption{The mean reconstruction error of the temporal algorithm for \textsl{LM-SG}.}
		\label{fig: 6}
\end{figure}
\subsection{The number of reporting sensors (Cost)}
Cost and performance are two competing factors that usually improving one, results in losing the other one. In this letter, by selecting a proper set of contour levels $\{\ell_i\}_{i=1}^M$ and their margin $\Delta$, we show that it is possible to simultaneously save cost and performance. The number of reporting sensors as cost, plays an important role is this algorithm. Fig.~\ref{fig: 3}, illustrates the cumulative spatial monitoring cost of the algorithm through the learning process. As this figure shows, by increasing the number of the levels, the spatial monitoring cost of \textsl{LM-SG} is around the cost of \textsl{U-SG}, where these costs are tangibly less than that of \textsl{LM-fixed}.

Fig.~\ref{fig: 4} illustrates the temporal monitoring cost, once the IFC queries the sensor field periodically. As this figure shows, almost steadily around 9\% to 10\% of the sensors are reporting to the IFC.
\subsection{The monitoring performance}
Fig.~\ref{fig: 5} and Fig.~\ref{fig: 6} illustrate the performance of the spatial and the temporal monitoring, respectively. As Fig.~\ref{fig: 5} illustrates, by increasing the number of contour levels the performance of all 3 schemes is improved. This figure also shows that the performance of \textsl{LM-SG} is between that of \textsl{U-SG} and \textsl{LM-fixed}, due to success in estimation of \textit{pdf}. Fig.~\ref{fig: 6}, illustrates the performance of \textsl{LM-SG} in periodic temporal updates. This figure shows that the performance of temporal monitoring closely swings in a few dBs around that of spatial monitoring.
\subsection{The stochastic gradient learning step }
\begin{figure}[h!]
		\centering
		\includegraphics[width=2.9in,angle=0]{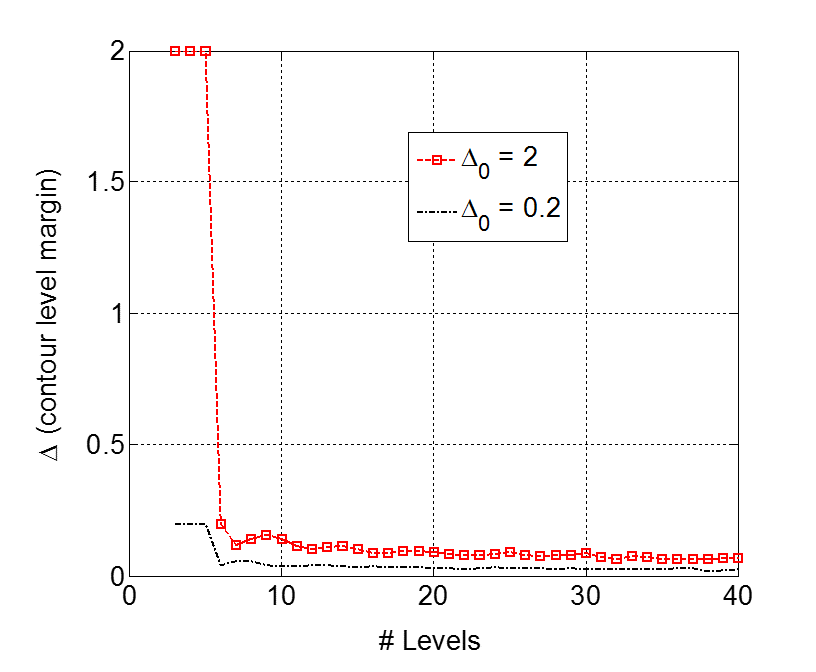}\\
		\caption{The variation of contour margin ($\Delta$) in \textsl{LM-SG}  (noise $\sigma$ = 0).}
		\label{fig: 7}
\end{figure}

The introduced stochastic gradient algorithm, manages the margin $\Delta$. According to Fig.~\ref{fig: 7}, $\Delta$ convergence to a tight bound for different initial $\Delta$. All results are related to Fig.~\ref{fig: 1}.
\section{Conclusion}
A cost efficient algorithm is presented and discussed for spatiotemporal monitoring of correlated signals in localized wireless sensor field. The signal distribution is compressed into its contour lines. The proposed algorithm uses a stochastic gradient approach to reduce the monitoring cost. The algorithm assumes no initial knowledge of the signal such as the signal strength, its statistical characteristics, its spatial, spectral or temporal attributes. The evaluation results show the steady convergence and the low cost attributes of the algorithm.



\end{document}